# NUMERICAL SIMULATION OF A DROP COLLISION WITH AN OBSTACLE

*A.I. Fedyushkin[#], A.N. Rozhkov[##]*

*Ishlinsky Institute for Problems in Mechanics RAS, Moscow, Russia*

[*]Address correspondence to: Alexey Fedyushkin, IPMech RAS, 101(1) Prospect Vernadskogo, Moscow 119526, Russia, E-mails: [#] fai@ipmnet.ru, [##] rozhkov@ipmnet.ru

**Abstract**

The paper considers the dynamics of the spreading liquid droplets after impact on a solid surface. The dynamics of a drop falling on a solid surface is shown using numerical simulation based on the solution of 2D and 3D Navier-Stokes equations for a two-layer liquid-gas system. The results of numerical simulation are compared with experimental data.

| NOMENCLATURE | |
|---|---|
| $d_i$ | impact drop diameter (m) |
| $d$ | rim diameter (m) |
| $d_m$, $d_{max}$ | maximum rim diameter (m) |
| $d_t$ | target diameter (m) |
| $\boldsymbol{F}$ | force vector (N) |
| $\boldsymbol{g}$ | acceleration vector of the Earth's gravity (m/s$^2$) |
| $g$ | acceleration of the Earth's gravity (m/s$^2$) |
| $h$ | height of interface (mm) |
| $H$ | height of calculation region (m) |
| $k$ | surface curvature (m$^{-1}$) |
| $l_s$ | point on interface |
| $\boldsymbol{n}$ | unit normal vector (m) |
| $p$ | pressure (Pa) |
| $r, z$ | radial and axial coordinates (m) |
| $R$ | radius of calculation region (m) |
| **u** | vector of fluid velocity (m/s) |
| $u_1$ | radial component of fluid velocity(m/s) |
| $u_2$ | axial component of fluid velocity(m/s) |
| $v_i$ | impact velocity (m/s) |
| $t$ | time (s) |
| $x, y, z$ | Cartesian coordinates (m) |
| $Re_i$ | impact Reynolds number |
| $We_i$ | impact Weber number |
| **Greek Symbols** | |
| $\beta_m$ | dimensionless diameter of the maximum spreading |
| $\varepsilon$ | function of the volume phase fraction |
| $\gamma$ | surface tension (N/m) |
| $\theta$ | wetting angle (degree) |
| $\mu$ | dynamic viscosity (Pa s) |
| $\rho$ | density (kg/m$^3$) |
| $\tau_{\alpha ij}$ | viscous stress tensor (Pa) |
| $\zeta_1, \zeta_2$ | radii of curvature (m) |
| **Subscripts** | |
| $\alpha$ | phase index |
| i | impact |
| m, max | maximum |
| s | point |
| **Superscripts** | |
| $T$ | transposed |
| | |

## 1. INTRODUCTION

The study of droplet spreading when falling on a solid surface has both fundamental scientific significances for the study of the laws of multiphase hydrodynamics, and applied significance in various physical processes, for example, in metallurgy, energy, micro-technologies, in the electronic, nuclear and aviation industries, medicine and health care, in the processes of cooling, irrigation, fire extinguishing, in jet and 3D printing and in other areas.

A lot of papers devoted to the study of droplet dynamics, for example, works, (Rozhkov et al., 2002, 2004, 2006, 2010, Rozhkov 2005, Fedyushkin and Rozhkov,2018, 2020, 2021,

Yarin et al, 2017, Yarin and Weiss, 1995). Reviews of works devoted to the study of hydrodynamics during impact and spreading of droplets can be found in (Yarin, 2006, Rozhkov, 2005). This paper presents the results of mathematical modeling of the spreading of water droplets falling on a solid surface based on the numerical solution of the Navier-Stokes equations for two-phase air-liquid systems. The simulation results are compared with experimental data.

## 2. NUMERICAL MODELLING TWO-PHASE AIR-LIQUID SYSTEM

### 2.1. Mathematical model

Simulation of the process drop collision with obstacles was compared with experiment which scheme shown in Fig. 1 (Rozhkov et al., 2004, 2006; Vernay, 2015; Vernay et al., 2015; Villermaux and Bossa, 2011). Numerical simulation is based on the numerical solution of three- and two-dimensional axisymmetric Navier-Stokes equations for two immiscible liquids (Landau and Lifshitz, 1987).

$$\frac{\partial(\rho \boldsymbol{u})}{\partial t} + \nabla \cdot (\rho \boldsymbol{u}\boldsymbol{u}) = -\nabla p + \nabla \cdot \left[\mu\left(\nabla \boldsymbol{u} + \nabla \boldsymbol{u}^T\right)\right] + \rho \boldsymbol{g} + \boldsymbol{F} \quad . \tag{1}$$

For two-phase air-liquid system the only one momentum equation is solved in all calculation domain, and the resulting velocity field is shared among the phases. The momentum equation (1) is dependent on the function of the volume fractions $\varepsilon$ ($0 \leq \varepsilon \leq 1$) of liquid phase through the properties: density $\rho = \varepsilon \rho_{air} + (1-\varepsilon)\rho_{liquid}$ and viscosity $\mu = \varepsilon \mu_{air} + (1-\varepsilon)\mu_{liquid}$, where the values with the index 'air' refer to air, and with the index 'liquid' refer to liquid. The volume fraction of liquid $\varepsilon$ was determined from the solution of the transfer equation: $\frac{\partial \rho \varepsilon}{\partial t} + \nabla \cdot (\rho \varepsilon \boldsymbol{u}) = 0$.

For the two-dimensional case of a two-phase incompressible air-liquid system the Navier-Stokes equations (3) in a cylindrical coordinate system for an axisymmetric approximation without taking into account the circumferential velocity can be written as follows:

$$\frac{\partial u_1}{\partial x_1} + \frac{u_1}{x_1} + \frac{\partial u_2}{\partial x_2} = 0, \qquad (2)$$

$$\frac{\partial(\rho u_1)}{\partial t} + u_1 \frac{\partial(\rho u_1)}{\partial x_1} + u_2 \frac{\partial(\rho u_1)}{\partial x_2} = -\frac{\partial p}{\partial x_1} + \frac{1}{x_1} \frac{\partial}{\partial x_1}\left(x_1 \mu \frac{\partial u_1}{\partial x_1}\right) + \frac{\partial}{\partial x_2}\left(\mu \frac{\partial u_2}{\partial x_2}\right) - \mu \frac{u_1}{x_1^2} + F_1, \qquad (3)$$

$$\frac{\partial(\rho u_2)}{\partial t} + u_1 \frac{\partial(\rho u_2)}{\partial x_1} + u_2 \frac{\partial(\rho u_2)}{\partial x_2} = -\frac{\partial p}{\partial x_2} + \frac{1}{x_1} \frac{\partial}{\partial x_1}\left(x_1 \mu \frac{\partial u_2}{\partial x_1}\right) + \frac{\partial}{\partial x_2}\left(\mu \frac{\partial u_2}{\partial x_2}\right) - \rho g + F_2, \qquad (4)$$

where $t$ is the time, $x_1 \equiv r$, $x_2 \equiv z$ are the radial and axial coordinates, $u_1$, $u_2$ are the radial and axial components of the velocity vector $\boldsymbol{u}(u_1, u_2)$, $p$ is the pressure, $\rho$ is the density, $\mu$ is the dynamic viscosity coefficient, and $F_1, F_2$ are the radial and axial components of external force $\boldsymbol{F}(F_1, F_2)$, acting only in very narrow zone along the air-liquid interface, $g$ is axial component of the gravity acceleration vector $\boldsymbol{g}$.

To describe the fluid flow of a two-phase air-liquid system in the approximation of axial symmetry, a model based on 2D system of Navier-Stokes equations (2–4) was used, which made it possible to determine the velocities for liquid and air. Density and viscosity were determined through the function of the volume fraction of the liquid ε as $\rho = \varepsilon \rho_{air} + (1-\varepsilon)\rho_{liquid}$, $\mu = \varepsilon \mu_{air} + (1-\varepsilon)\mu_{liquid}$. The function of the volume fraction of the liquid $\varepsilon$ was determined from the solution of the transfer equation: $\frac{\partial \varepsilon}{\partial t} + u_1 \frac{\partial \varepsilon}{\partial x_1} + u_2 \frac{\partial \varepsilon}{\partial x_2} = 0$.

The following boundary conditions were determinate: at the remote external boundaries - flow symmetry conditions, on a solid wall, there were boundary conditions no slip with a wetting angle of $\theta$, at the air–liquid interface were determined from the equilibrium condition of surface tension forces and pressure forces (Landau and Lifshitz, 1987):

$$(p_1 - p_2 + \gamma k)n_i = (\tau_{1ij} - \tau_{2ij})n_j \tag{5}$$

where $\gamma$ is the surface tension coefficient assumed to be constant; $p_1$, $p_2$ are the liquid and air pressures; $k = 1/\zeta_1 + 1/\zeta_2$ is the surface curvature, where $\zeta_1$, $\zeta_2$ are the radii of curvature of the air–liquid interface; $\boldsymbol{n}$ is the unit normal vector at the interface directed from the liquid into the air; $\tau_{\alpha ij}$ is the viscous stress tensor and phase index $\alpha$ denotes: $\alpha = 1$ – liquid, $\alpha = 2$ – air. Condition (5) is written for a constant the surface tension coefficient $\gamma$. In the case of variability of surface tension (for example, these are cases of heat- or concentration-capillary convection) necessary additionally to take into account the changing of the surface tension force along the interface ($l$), and append the term $\frac{\partial \gamma}{\partial l}$ to the right side of the tangential projection of expression (5).

Changing the interphase boundary shape of the air-liquid interface was performed using the volume of fluid method (VOF). The interface was defined using an implicit VOF method with Euler iterations. To increase the accuracy of calculating the location of the interface, a piecewise linear geo-reconstruction approach was used (Youngs et al., 1982). The interphase boundary was determined from the solution of the system of equations (2–6) taking into account surface forces by the method of continuous surface forces (CSF) (Brackbill et al., 1992). The CSF method allows the condition (7) at the interface to be taken into account through an additional local bulk force $\boldsymbol{F}$ in the right part of the momentum transfer equation (5–6). The force $\boldsymbol{F}$ acts only in a very narrow zone enclosed along an interface line of width $\Delta h$ (Fig. 2). For every point $l_s$ of interface lines $l$, at $\Delta h$ striving to zero, the force $\boldsymbol{F}$ can be written, for example, as $\boldsymbol{F}(l_s) = \gamma k(l_s)\boldsymbol{n}(l_s)$ that was shown by Brackbill et al., 1992, where $k(l_s)$ is the interface curvature at a point $l_s$, and $\boldsymbol{n}(l_s)$ is the normal to the site at the interface point $l_s$ (Fig. 2).

The CSF method makes it possible to eliminate the singularity in the case of turning the radius of curvature of the interface to zero and improve the accuracy of calculations when calculating forces at the interface in the form $\boldsymbol{F} = 2\gamma\rho k\nabla\varepsilon/(\rho_{air}+\rho_{liquid})$ (Brackbill et al., 1992) .For numerically solve the Navier-Stokes equations the conservative control volume method (Patankar, 1980) was used with an adaptive grid. To verify this VOF-CSF mathematical model and determine the accuracy of reproducing changes in the shape of the interface between two liquids, the simulation results were compared with spreading experiments (Rozhkov et al., 2004; Fedyushkin and Rozhkov, 2018) and droplet coalescence (Fedyushkin and Rozhkov, 2014, 2020). In addition, the dynamic problem of changing the interface level of a two-layer system in time, rotating rapidly in a cylinder, was solved and compared with experimental data published in the work (Sugimoto and Iguchi, 2002). When modeling this test problem, the presence of a circumferential velocity was taken into account and the system of equations (2–4) was supplemented by the equation of its momentum transfer. Comparison of the simulation results of the test problem with experimental data (Sugimoto and Iguchi, 2002) showed good accuracy and are given in the Appendix.

## 3. Results of numerical simulation

The calculation areas and grids for 3D (a) and 2D (b, c, d) models are shown in Fig. 3. The dimensions of 2D regions were height $H$ and width $R$, and in the case of a 3D model, the geometry of the calculation area was a truncated cone with height $H$ with radii of the lower circle $R$ and upper $H/2$. To exclude the influence of external boundaries, the dimensions of the calculated areas ($H$ and $R$) were much larger than the initial diameter of the drop ($d_i$), for example, in the 2D model 5 <$R/d_i$ <25. The sizes of the "target" on which the drop falls were different, but always larger than the diameter of the drop $d_t > d_i$, the same as in the experiments. Some series of calculations, for example, to determine the effect of viscosity and wetting angle, were performed for a "target" - horizontal bottom flat of all calculation region with size of $d_t/2 = R$. Non-uniform and dynamic grids were used in 2D and 3D models. For the 3D model, along with hexagonal and triangular grids, polygonal grids were used to reduce the calculation time (Fig. 3 (a)). Dynamic grids were used to reduce the calculation time, for example, in the case of 2D shown in Fig.3 d, the sizes of the calculation area $H$ and $R$ varied over time according to changes the shape of the drop. In the mathematical model, it is assumed that a drop at the moment of contact with an obstacle has a spherical shape with a diameter of $d_i$ and a velocity of $v_i$. A drop after impact on solid surface deformation into a thin lamella which can elongate, fragment and change into a new drop due to surface tension forces (Fig. 6, 8, 9).

### 3.1. 3D simulation of drop spreading after impact on plane surface

In Fig. 4 experimental data on the spreading of a water droplet and the results of 3D modeling of the dynamics of a drop of water ($v_i$=3.87 m/s, $d_i$=1 mm) falling on a solid hydrophobic surface showing the formation of a lamella, an edge jet and the separation of secondary droplets aproximately after $t$=1ms of impact on a solid surface are presented. In Fig. 4 (a) are shown side view of a drop: in left it is photo - experiment and right 3D modeling result as cross section of drop by plane y=0. In Fig. 4 (b) are shown top view of a drop: left photo - experiment and right - results of 3D simulation isolines of the water fraction and the rightmost picture - isolines of the velocity module from which the contour of a spreading drop where an edge jet is visible for the dimensionless time $\tau=t/(d_i/v_i)$ approximately equal to 2 (Fig. 4).

Depending on the time of change of the maximum spreading ($R_m = d_m/2$) drops with a diameter $d_i = 1$ mm falling on a solid surface at a velocity $v_i = 3.87$ m/s to the origin of coordinates are shown in Fig. 5. Lines 1 and 2 show the change on time of the maximum spreading distances on the plane $x=0$ in both directions from the origin of coordinates along the y direction, lines 3 and 4 are the dependencies of $R_m$ on time on the plane y=0 along both directions from the origin of coordinates along the x direction. Line 5 in Fig. 5 shows the change in the maximum thickness of the spreading drop (film) - the height of the spreading drop on the z axis. These dependencies show the features of the spreading of the drop, the time and asymmetry degree of the drop spreading. In the work (Rozhkov et al., 2004) it was shown that these dependences for dimensionless variables are universal and weakly depends on the properties of the liquid (Fig.11-13).

### 3.2. Initial stage of impact drops on solid surface

The results of experiments and mathematical modeling have shown that the initial stage of interaction of a falling drop with a solid surface is short-term and non-trivial. In Fig. 6 the dynamics of a falling water drop in the vicinity of the contact of the drop with the target and the formation of an initial jet and a lamella with an edge jet at the initial moments of interaction with a solid surface are shown. Firstly, when a drop falls at high velocity and collides with a solid surface, it is possible to capture air (in the form of a bubble or a mini torus) between the drop and the surface (Weiss and Yarin, 1999). Secondly, in initial moment of contact drop with the surface, a very thin and, as a result, very fast initial jet is formed (Fig. 6 (a - d)), and then a lamella with an edge jet is formed (Fig. 6 (e)), which schematically was depicted also in Fig. 1. In Fig. 6 (f) the results of the formation of an edge jet at the initial moment of impact of a drop on a hydrophobic solid surface with a wetting angle of $\theta = 180^0$ ($t = 6.8 \cdot 10^{-2}$ ms), which show the separation of the jet from the surface in contrast to the case of the jet flow on a hydrophilic surface (Fig. 6 (a–e)) are shown.

**3.3. A drop spreading**

The dynamics of the spreading of a falling drop depends not only on its diameter and rate, but also on the properties of the surface on which it falls and on which it spreads.

In Fig. 7 the numerical calculated dependences of the maximum spreading diameter of the lamella $d_m$ on time are shown. In Fig. 7a the spreading diameter for a frictionless surface with θ = 180º for small drop size ( $d_i$=2.67 mm, $v_i$=3.87 м/сек, $We_i$=550) and for large drop ($d_i$=3.89 mm, $v_i$=3.84 m/sec, $We_i$ =792) are presented . In Fig. 7b effect of droplet spreading the ($d_i$ =3.89 mm, $v_i$ =3.84 m/sec) on the surface with and without friction for wetting angel θ = 135º was shown. Numerical simulation of the spreading of a falling drop ($d_i$ =3.89 mm, $v_i$ = 3.87 m/s) on surfaces with different wetting angles was carried out. In Fig. 7c the dependences of the droplet spreading diameter for surfaces with and without friction for different wetting angles (θ = 0º, 90º, 135º and 180 º) are shown.

In the paper (Rozhkov et al., 2006), the specificity and difference of the dynamics of the spreading of "large" and "small" drops was shown, therefore, mathematical modeling of the spreading of a "large" drop ($d_i$ = 3.89 mm, $v_i$ = 3.84m/s) after its impact on the target ($d_t$ = 4 mm) was carried out, corresponding to the case of relatively large Weber ($We_i$ = 792) and Reynolds ( $Re_i = \dfrac{v_i d_i \rho}{\mu} = 1.5 \cdot 10^3$ ) numbers. The results of numerical simulation are presented in Fig. 8 shows the dynamics of changing the shape of a drop of the distribution of water fractions during the fall and spreading of a drop on a solid surface ($d_i$= 3.89 mm, $v_i$=3.83 m/s, $We_i$ = 792, $Re_i$ = 1.5 $10^3$) for different time moments in range from $t$ = 0.007 ms until $t$ = 2.23 ms. The results of experiments and mathematical modeling have shown that over time, due to the instability of the liquid-air interface, small droplets break off from the lamella, as shown in Fig. 1 and Fig. 8. Mathematical modeling for this case was conducted up to the time $t$ = 10 ms. After reaching the maximum spreading of the dm droplet, the remainder of the lamella due to the tension forces begins to contract ($t = 5.5$ ms) back to the target, collapses and breaks away from the target in the vertical direction ($t = 8.5$ ms), and later falls down on the target in the second time (Fig. 9).

### 3.4. The tightening of lamella

When the droplet spreads (at high Reynolds and Weber numbers), there comes a moment when the forces of viscosity and surface tension balance the forces of inertia (the expansion rate of the lamella slows down and the diameter of the edge jet of the lamella reaches the maximum value of $d_m$), the expansion of the lamella stops and it begins to tighten under the prevailing surface tension forces.

In Fig. 9 shows a continuation of the fall drop process shown in Fig. 8 of the collapse of the lamella formed during fall a drop on a solid surface (this is ($d_i$=3.89 мм, $v_i$=3.84 м/сек, We=792). In Fig. 9 the dynamics of tightening lamellar, starting from the time when the returning edge jet of the lamella has already reached the edge of the "target", i.e. when the diameter of the lamella is less than the diameter of the "target" " ($d_t$=4мм) are shown. The time from the beginning of the drop falling (touching the surface) is indicated in the figures. This case was considered without taking into account gravity (g= 0), therefore, after the collapse of the lamella, the newly formed drop rushes up and leaves the calculated region.

In Fig. 10 the dependences of the dimensionless diameter of the spreading droplet $\beta_m = d_m/d_i$ on the Weber number We in comparison with experimental data are shown. In Fig. 10 by line 1 - numerical results, line 2 - experimental data, line 3 - correlation expression for the maximum spreading of the droplet $\beta_m = (We_i/20)^{1/2}$ are presented. The results in Fig. 10 show a fairly good agreement with the experiment of the presented data, up to the value of the Weber number equal to 600.

### 3.5. Universal dependencies of droplet spreading

In paper Rozhkov et al., 2004 was shown that the spreading of the droplet is universal with respect to the dimensionless velocity $V=v/v_i$ and the dimensionless flow rate $Q=q/(\pi d_i^2 v_i/6)$, as a function of the dimensionless coordinate $Y=r/d_i$ and the dimensionless time $\tau=t/(d_i/v_i)$, respectively; $t$-time, $r$ is the radial coordinate (Fig. 1), $v_i$, the local velocity, and $q$ is the local flow velocity, is defined as the amount of liquid flowing through a circular contour of radius r per unit time, that is, $q=2\pi rhv$, where $h$ is the local thickness of the film. In Fig 11. Influence of conditions ($v_i$, $d_i$ liquid density $\rho$ and surface tension $\gamma$) shown in Fig. 11.

All curves can be approximated by universal functions $V(\tau,Y)$ and $Q(\tau,Y)$, characterizing the outflow of the liquid film from the point source in origin coordinates.

The results of the dependences of the dimensionless velocity $V=v/v_i$ and the dimensionless flow $Q=q/(\pi d_i^2 v_i/6)$ on the dimensionless coordinate $Y=r/d_i$ and the dimensionless time $\tau=t/(d_i/v_i)$ show that all curves form surfaces with a certain accuracy. These surfaces $V(\tau, Y)$ and $Q(\tau,Y)$ are a universal approximation and are shown in Figures 12a, and 12b. The results of numerical simulation in Fig. 11-12 presented by the lines. Figure 13 shows the distribution surface of the dimensionless film thickness $H=h/d_i$ depending on the dimensionless coordinate $Y$ and the dimensionless time $\tau$.

In Fig. 14 shown the dependences of droplet spreading diameters on time (a) – in dimensionless variables $d_i$=2.79 mm, $v_i$=3.40 m/sec, $We_i$=448, (b) – in dimensional variables $d_i$ =2.67 mm, $v_i$ =3.87 m/sec, $We_i$ =550 ("Small" drop) and $d_i$=3.89 mm, $v_i$ =3.84 m/sec, $We_i$ = 792 ("Large" drop). In Fig. 14, the experimental data are indicated by circles (O) (Rozhkov et al., 2004). Solid lines (E) are theoretical predictions based on universal outflow functions calculated on the basis of experimental data (Rozhkov et al., 2004). Dotted lines

(N) are theoretical predictions based on universal expiration functions calculated based on numerical simulation data.

## 4. CONCLUSIONS

The dynamics of a drop falling on a solid surface is shown using numerical simulation based on the solution of 2D and 3D Navier-Stokes equations for a two-layer liquid-gas system. The results of numerical simulation are compared with experimental data. The results of numerical modeling are generally in good agreement with experimental data and confirm the reliability of the developed approach.

Experimental and numerical results confirming universal patterns of droplet spreading on a solid surface are shown.

## ACKNOWLEDGMENTS

This work was supported by the Government program (contract # AAAA-A20-120011690131-7) and was funded by RFBR, project number 20-04-60128.

**FIGURES**

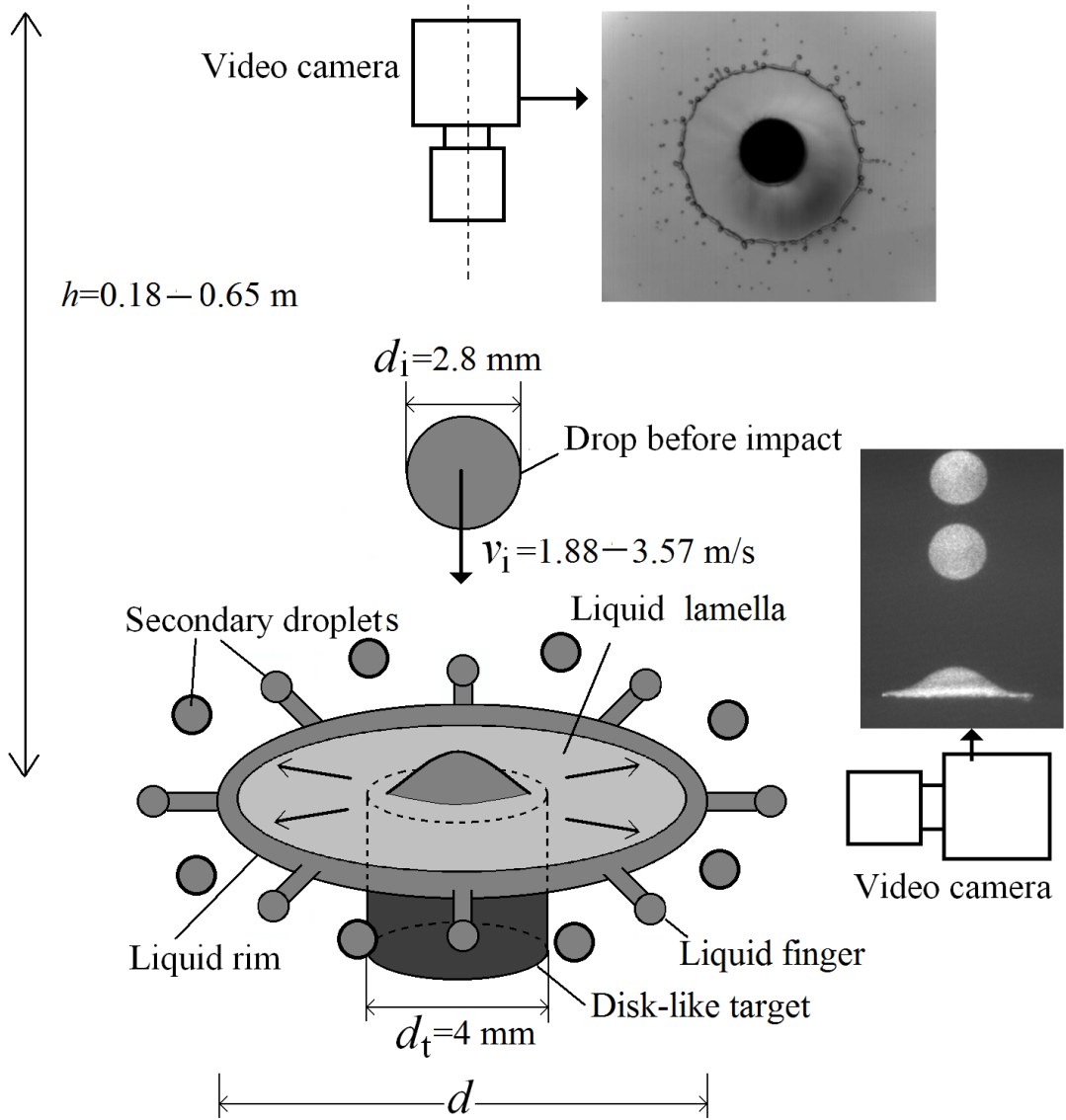

**FIG. 1:** Drop impact on small disc-like target

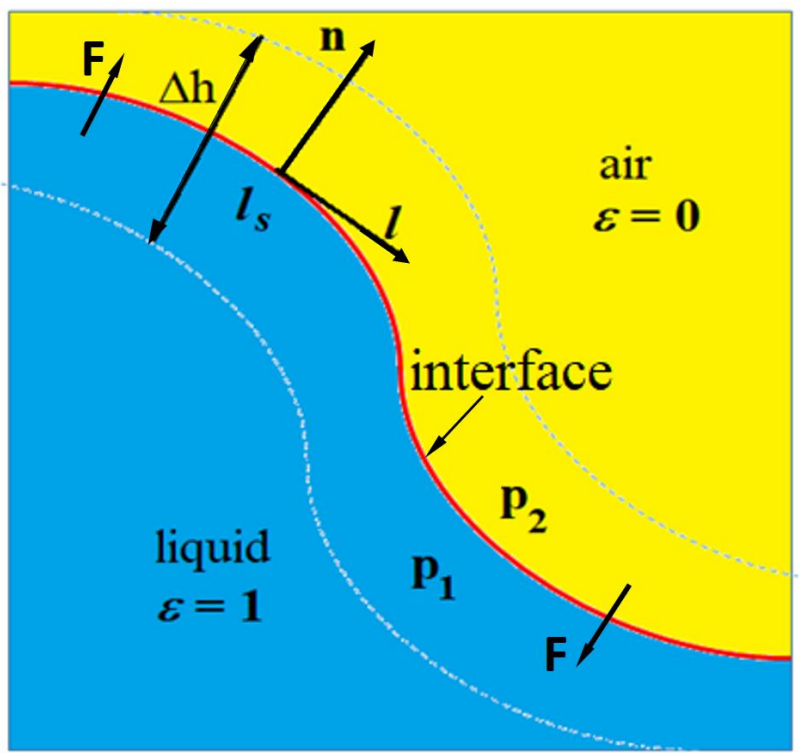

**FIG. 2:** The scheme of numerical method for definition of interface (Brackbill et al., 1992)

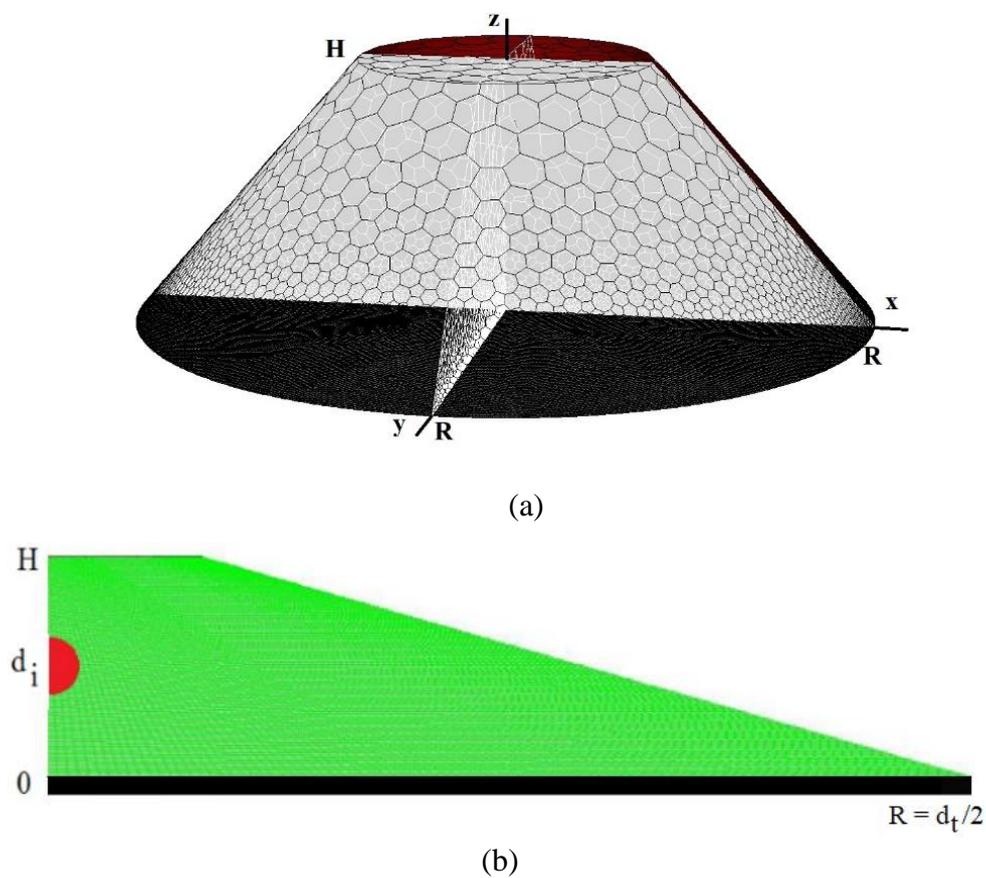

(a)

(b)

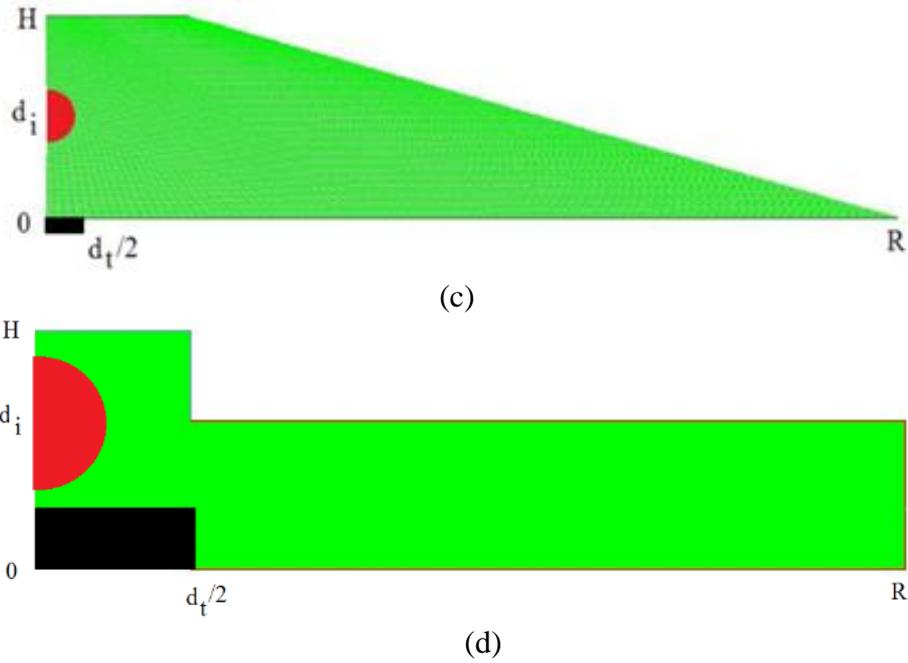

(c)

(d)

**FIG. 3:** Schemes of computational regions with grids for 3D (a) and 2D (b, c) models

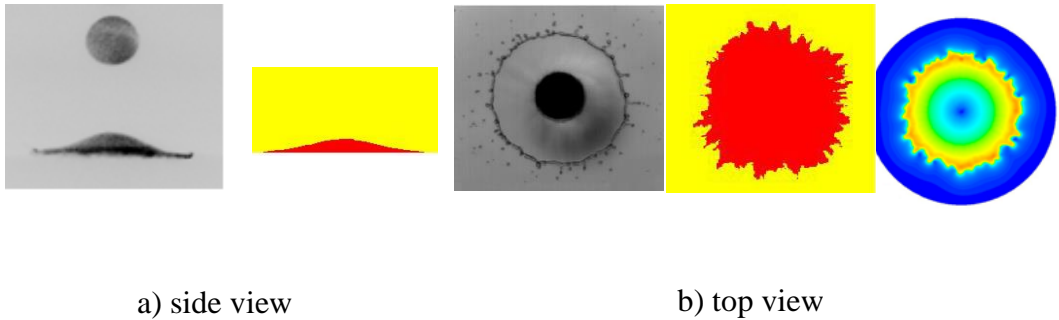

a) side view b) top view

**FIG. 4:** Spreading of water drop ($v_i$ = 3.87 m/s, $d_i$ = 1 mm) aproximately after $t$ = 1ms of impact on a solid surface ((a) - side view, (b) - top view), left photos - experiment, right - results of 3D modeling: isolines of the water fraction, the rightmost picture - isolines of the velocity module.

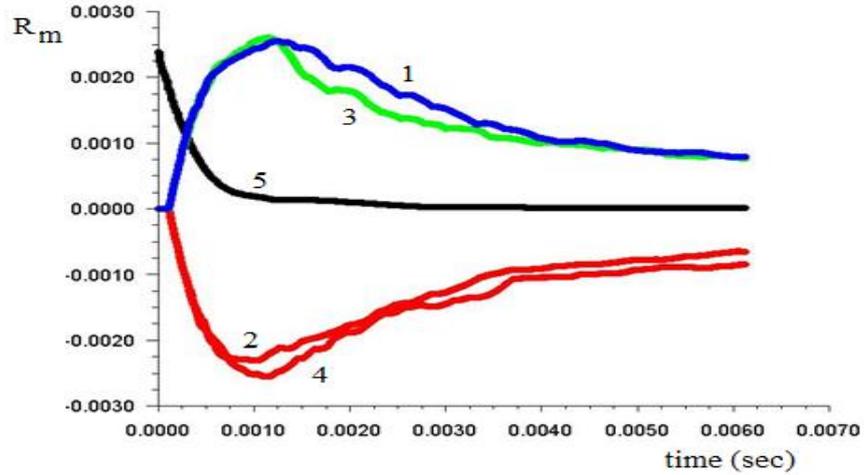

**FIG. 5:** The dependences of the change in the maximum spreading distance of the drop ($R_m$) on time along the main coordinate planes: lines 1, 2 are along the plane x=0; lines 3, 4 are along the plane y= 0; line 5 is the maximum thickness of the spreading drop (lamella) along the z axis).

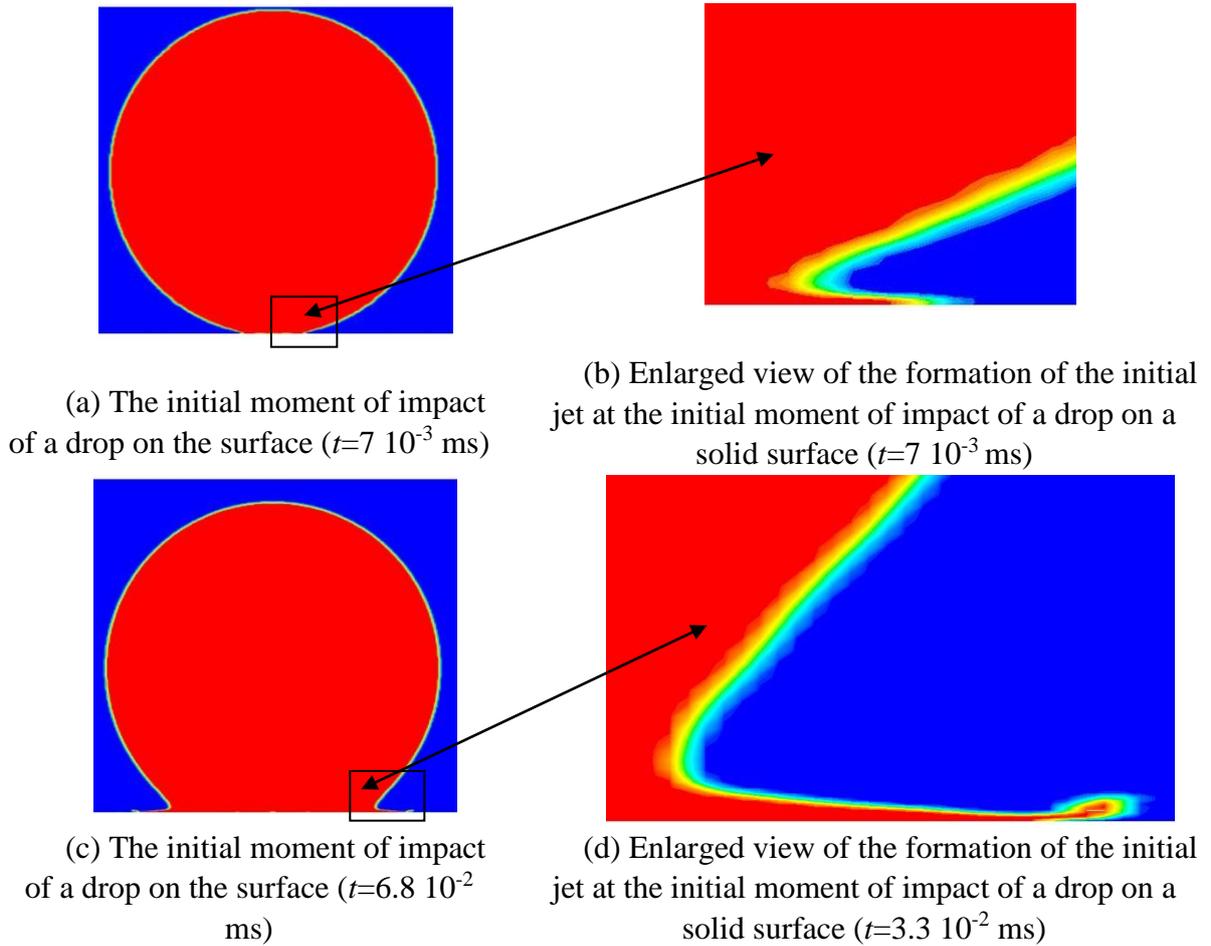

(a) The initial moment of impact of a drop on the surface ($t=7\ 10^{-3}$ ms)

(b) Enlarged view of the formation of the initial jet at the initial moment of impact of a drop on a solid surface ($t=7\ 10^{-3}$ ms)

(c) The initial moment of impact of a drop on the surface ($t=6.8\ 10^{-2}$ ms)

(d) Enlarged view of the formation of the initial jet at the initial moment of impact of a drop on a solid surface ($t=3.3\ 10^{-2}$ ms)

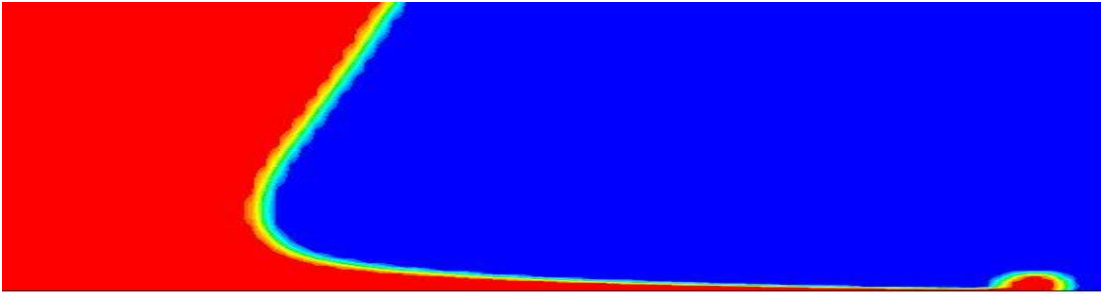

(e) Enlarged view of the formation of the lamella and the edge jet at impact of the drop on a solid surface ($t=0.1$ ms)

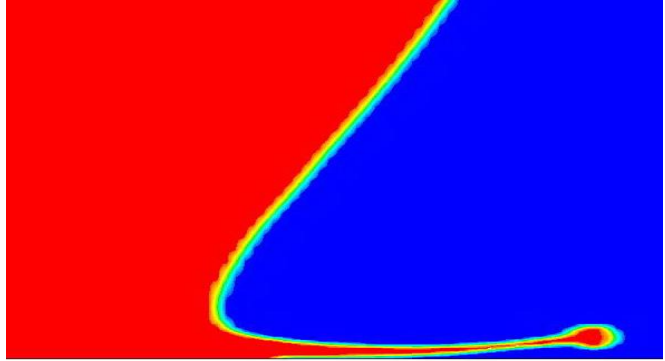

(f) Enlarged view of the formation of the initial jet at the initial moment of impact of a drop on a hydrophobic solid surface with wetting angle $\theta = 180^0$ ($t = 6.8 \cdot 10^{-2}$ ms)

**FIG. 6:** Formation of an initial jet and a lamella rim with an edge jet in initial moments of impact of a drop on a solid surface ($d_i = 2.67$ mm, $v_i = 3.87$ m/s)

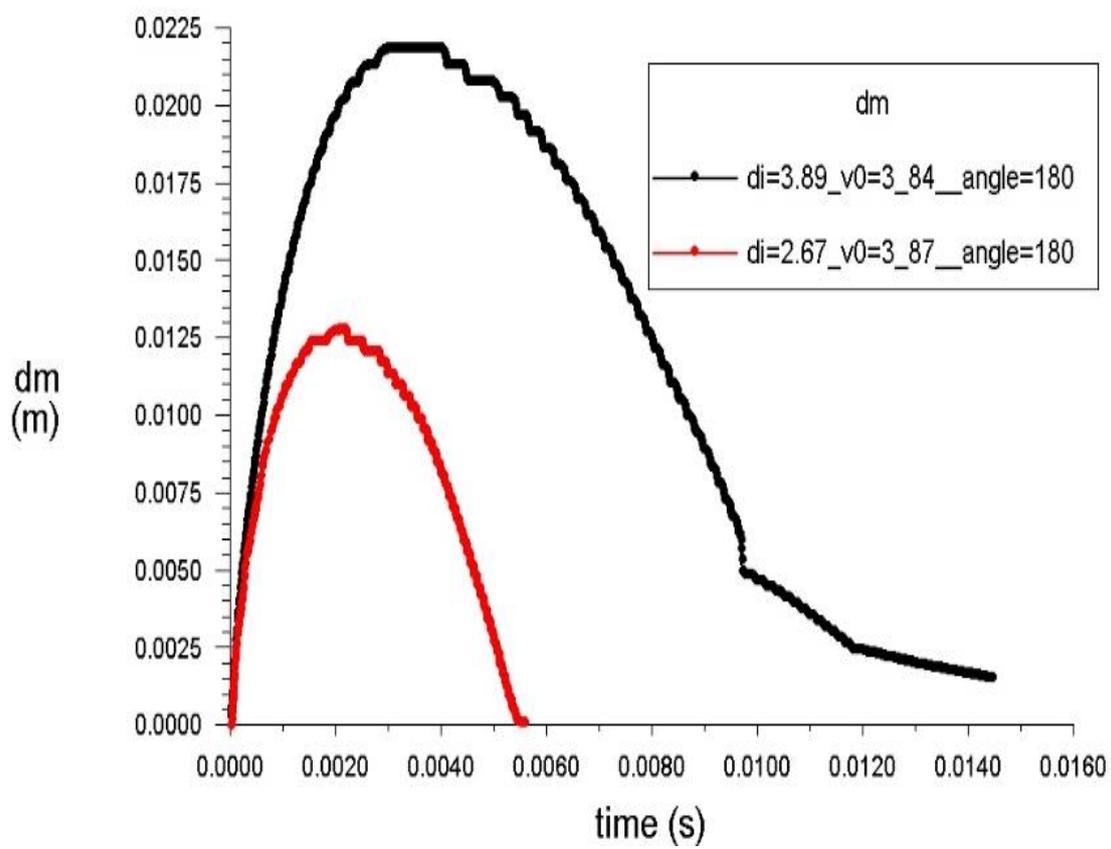

a)

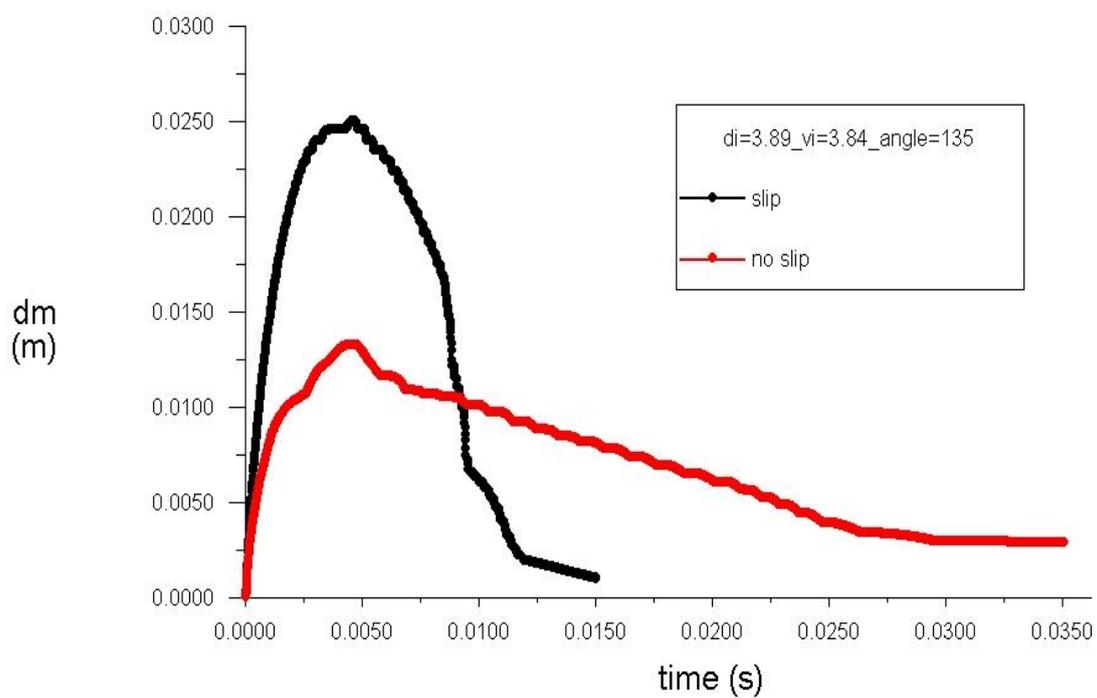

b)

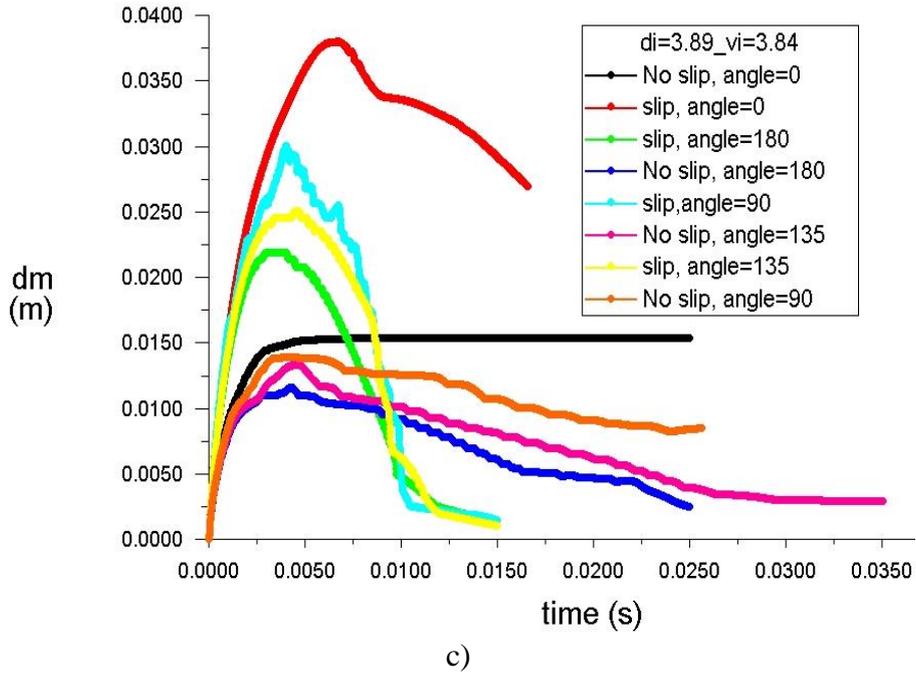

c)

FIG. 7: The dependence of the diameter of the spreading of the drop $d_m$ on time.

a) spreading of small drops ($d_i$ = 2.67 mm) and large drops ($d_i$ = 3.89 mm), b) spreading on surfaces with and without friction, c) spreading of drops for surfaces with and without friction for different wetting angles ($\theta$ = 0°, 90°, 135° and 180°).

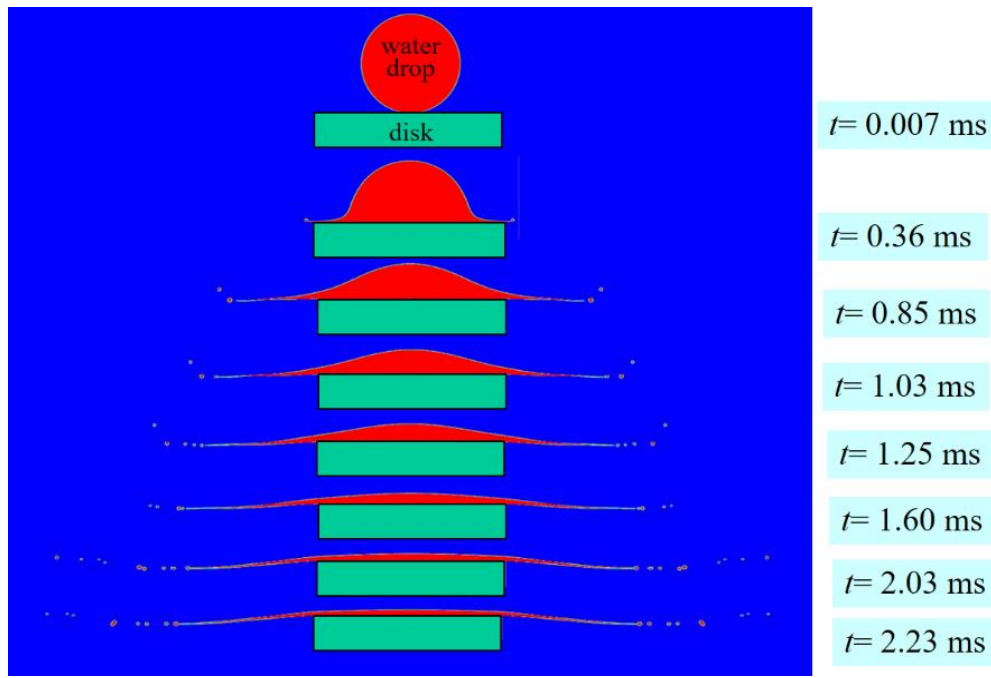

FIG. 8: Spreading of water drop ($d_i$= 3.89 mm, $v_i$= 3.84 m/s , $We_i$ = 792, $Re_i$ = 1.5 10$^3$) on a solid disk target for time from $t$= 0.007 ms until $t$= 2.23 ms

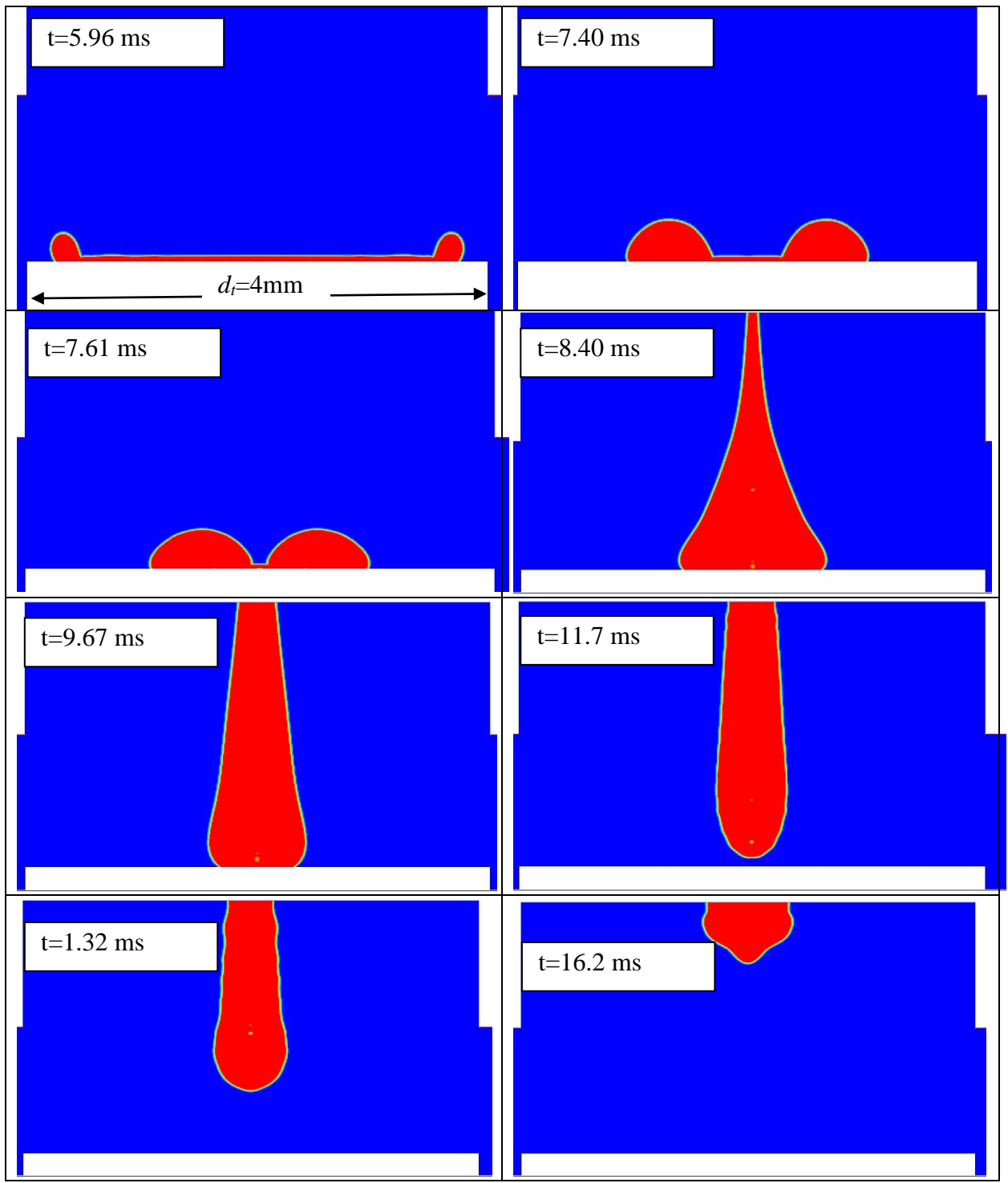

**FIG. 9:** The collapsing of the lamella formed during fall a drop on a hard surface (this is a continuation of the fall drop process shown in Fig. 8)

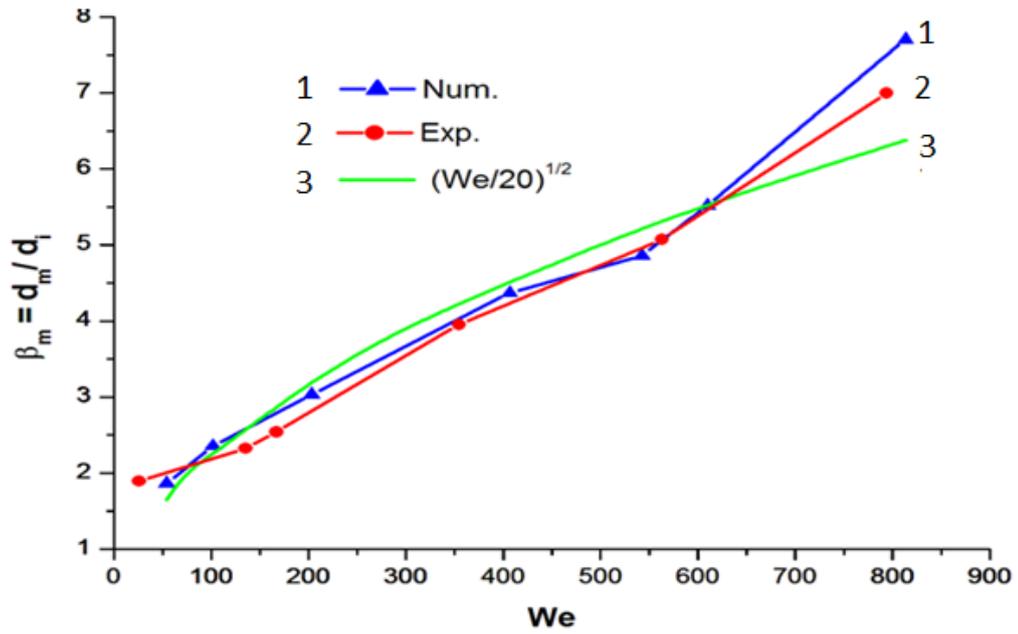

**FIG. 10:** Dependences of the dimensionless diameter of the maximum spreading of the droplet $\beta_m = d_m/d_i$ on the Weber number $We$ (line 1 - numerical results, 2 - experimental data, line 3 - approximation $\beta_m = (We_i/20)^{1/2}$)

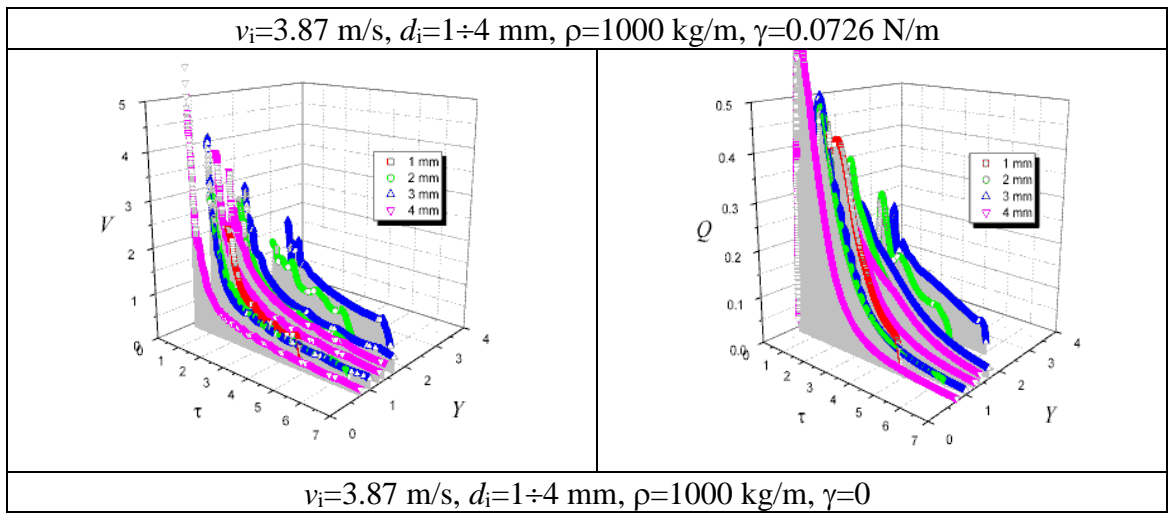

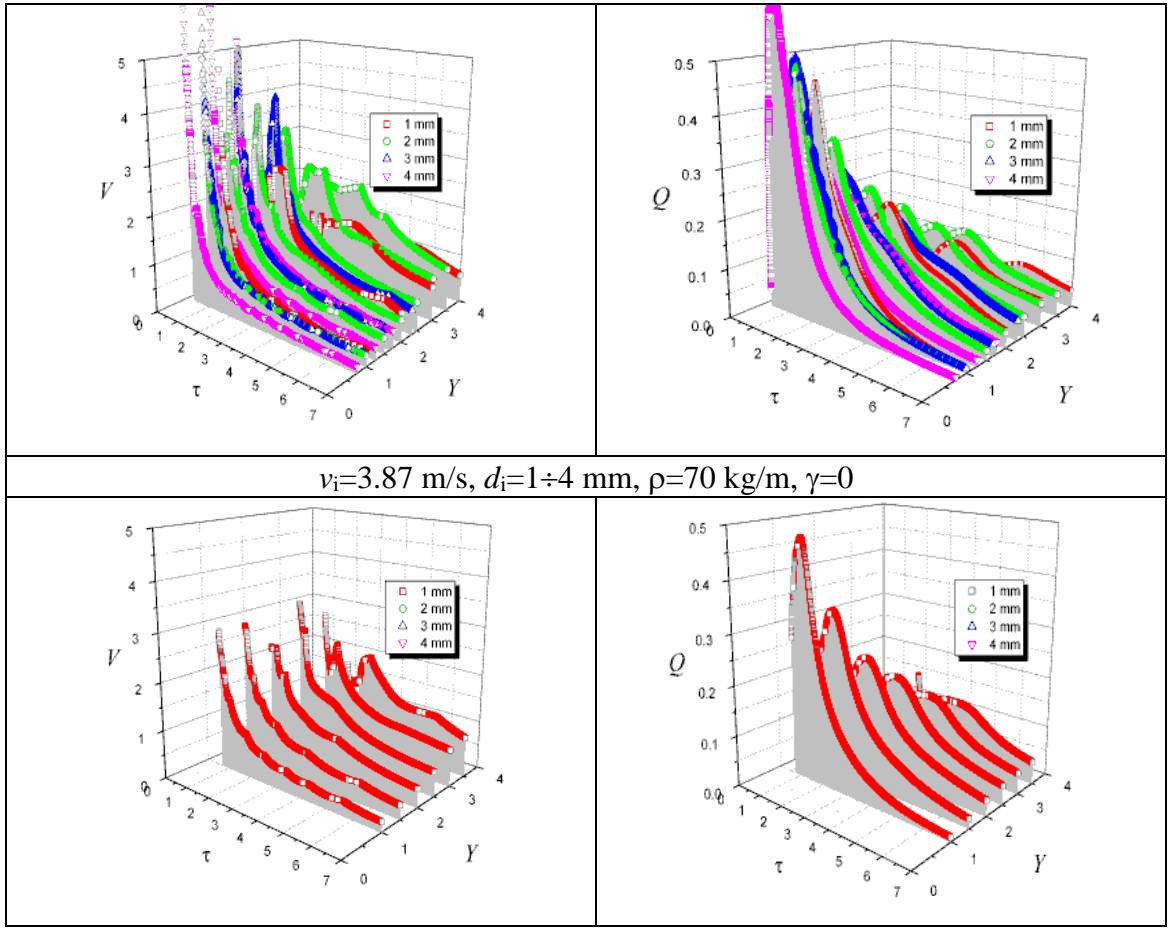

**FIG 11:** Universality of the flow in the impacting drop (it is resulted on numerical data): the dimensionless velocity V=v/vi and dimensionless flow rate Q=q/($\pi d_i^2 v_i$/6), as functions of dimensionless coordinate Y=r/$d_i$ and dimensionless time $\tau$=t/($d_i$/$v_i$), respectively; *t* is the time, *r*, the radial coordinate (Fig. 1), *v*, the local velocity, and *q*, the local flow rate defined as the amount of liquid that flows through a circular contour of radius *r* per unit time, i.e., $q=2\pi rhv$ with *h* the local film thickness.

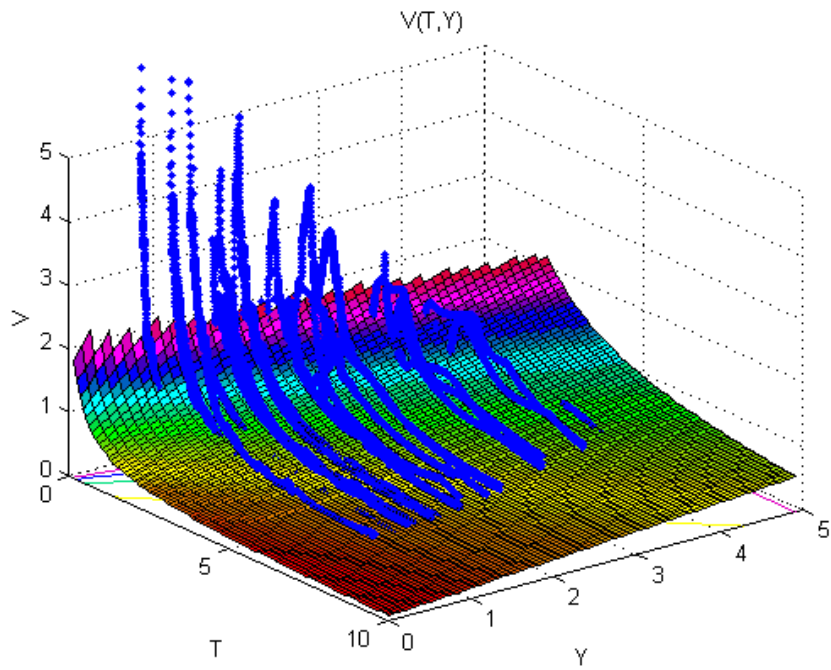

a)

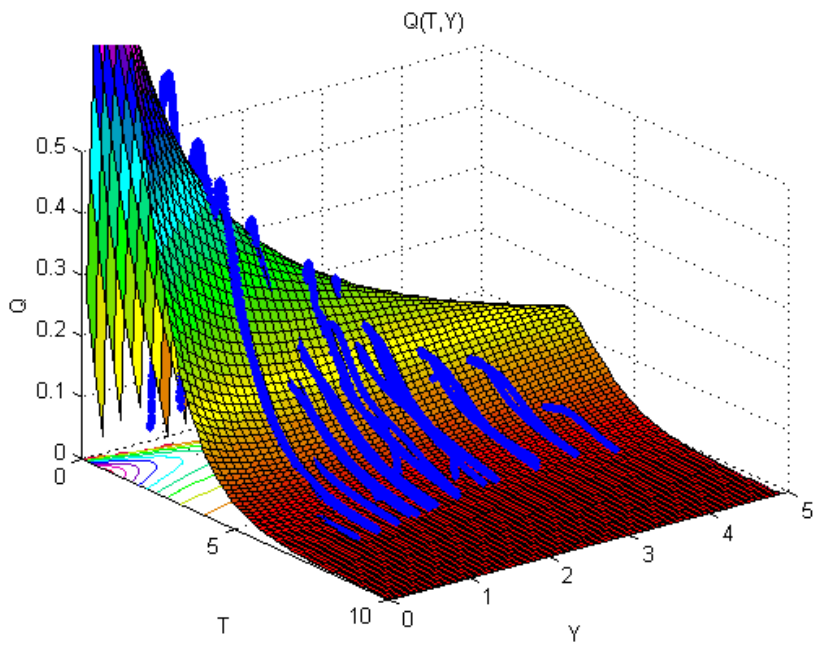

b)

**FIG 12:** Surfaces of universal functions (lines - calculation) a) - $V(\tau,Y)$, b) - $Q(\tau,Y)$.

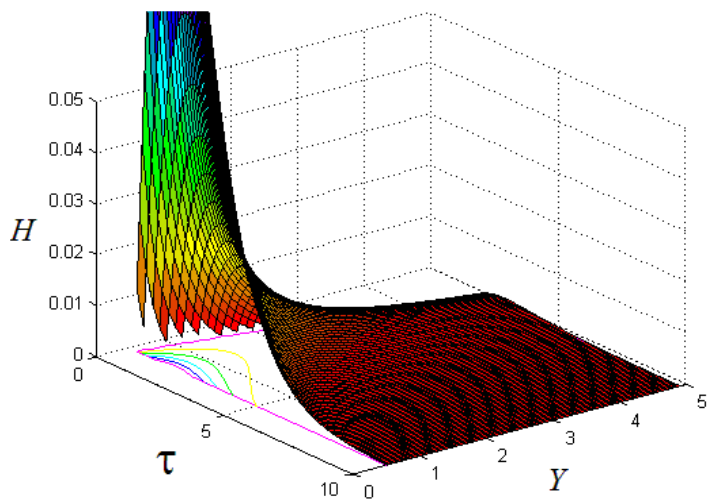

**FIG 13:** The surface of a universal function of dimensionless lamella thickness $H=h/d_i$ as a function of dimensionless time $\tau$ and dimensionless coordinate $Y$.

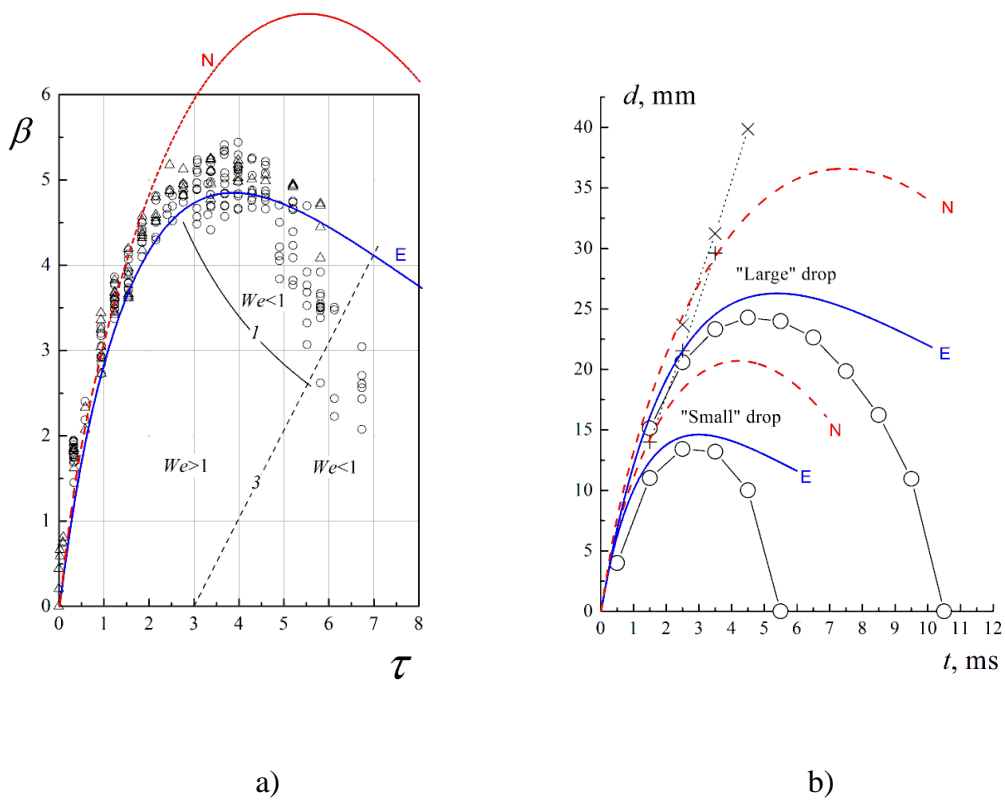

a)          b)

**FIG 14:** The dependences diameter of the spreading of droplets on time

a) – dimensionless variables, b) – dimensional variables

**Appendix**

# Simulation of changing in the interface level of a two-layer oil-water system rotating in a cylinder

The purpose of this test problem is to determine the possibilities of reproducing the unsteady effects of changing the shape of the interface of two-layer liquid systems in a cylinder by numerical simulation. The simulation is based on the numerical solution of the Navier-Stokes equations (2-5) with defining interface by VOF method. The numerical simulation results were compared with experimental data (Sugimoto and Iguchi, 2002).

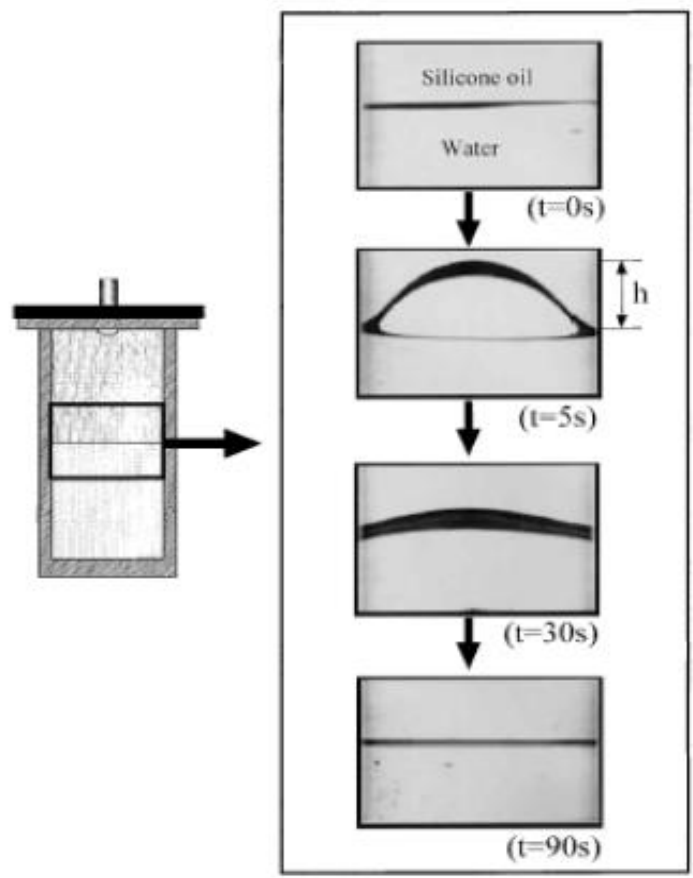

**FIG. 1:** Scheme of the experimental setup (left) and changing the shape of the interface between silicone oil and water (right) for different time moments ($v_{so}$=100 mm$^2$/s, $V_w/V_{so}$=1.0, $Re_w$=74.9). (The drawing is taken from paper of Sugimoto and Iguchi, (2002)).

**Test problem statement**

The calculation area is a vertical cylindrical vessel filled with immiscible liquids (silicone oil and water) which are at rest in the initial moment. There is a layer of silicone oil above the water. Oil and water have similar densities, but the viscosity of oil is 10 times greater than that of water: the viscosity of $\mu_{oil} = 0.0103$ [kg/m s], the density of $\rho_{oil}= 935$ [kg/m³] is higher than that of water $\rho_w= 1030$ [kg/m³], $\mu_w =0.00103$ [kg/m s]. The cylindrical vessel is covered with a lid from above. The cylinder suddenly begins rotation around its axis at a constant angular velocity. The diameter of the vessel is $2R = 46$ mm, and the height is $H_{so}+H_w=120$ mm, respectively (Fig.2).

Dimensionless parameters are defined as follows: $Re_\omega = R(\frac{\omega}{\nu})^{1/2}$ - Reynolds number, $R_V = \frac{V_w}{V_{so}}$ - ratio of volumes occupied by silicone oil and water. This paper presents the results of calculations for $R_v=1$. To model this test problem, a two-dimensional axisymmetric model (2-4) was used with an additional equation for the transfer of the circumferential velocity momentum. For numerically solve the Navier-Stokes equations the conservative control volume method (Patankar, 1980) with VOF method for defining of interface was used. It should be noted that the using of the VOF method imposes additional restrictions on the values of spatial and especially time steps. For the calculation, quadrangular grids with a different number of cells with a thickening near the interface were used.

**Numerical simulation results**

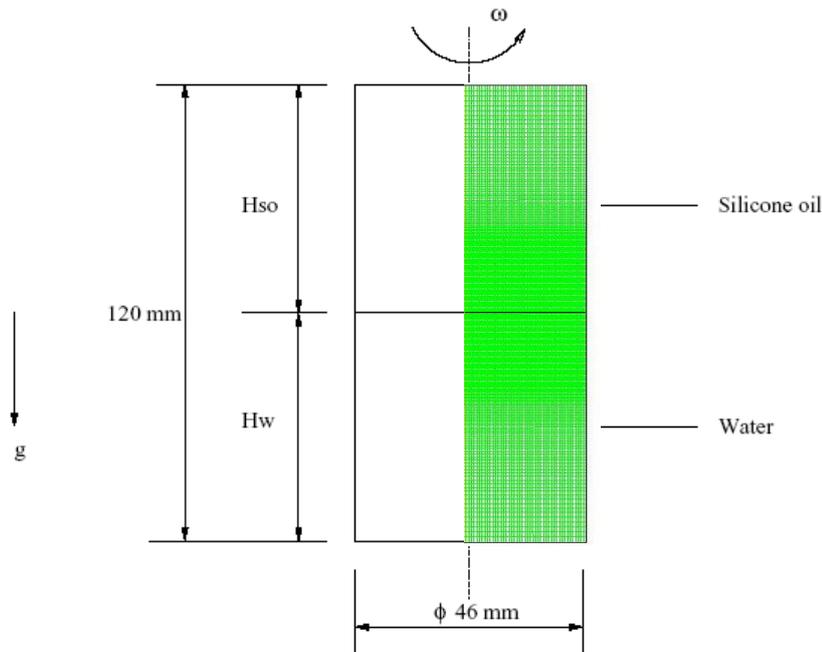

**FIG. 2:** Cylindrical calculation area and grid

When the cylinder filled with oil and water is instantaneously brought into rotational motion, due to the difference in viscosity of oil and water and as a consequence of different scales of dynamic times, a change in the interface will occur. At the initial moment, silicone oil (as more viscous) will begin to displace water at the cylinder wall, and give way to water on the cylinder axis. Changes in the magnitude and direction of the pressure forces near the walls of the cylinder leads to a change in the direction of the vector of pressure forces in the water at the axis of the cylinder, where the acting forces add up, and therefore the maximum deviation of the interface from its initial equilibrium position can be expected on the axis of the cylinder.

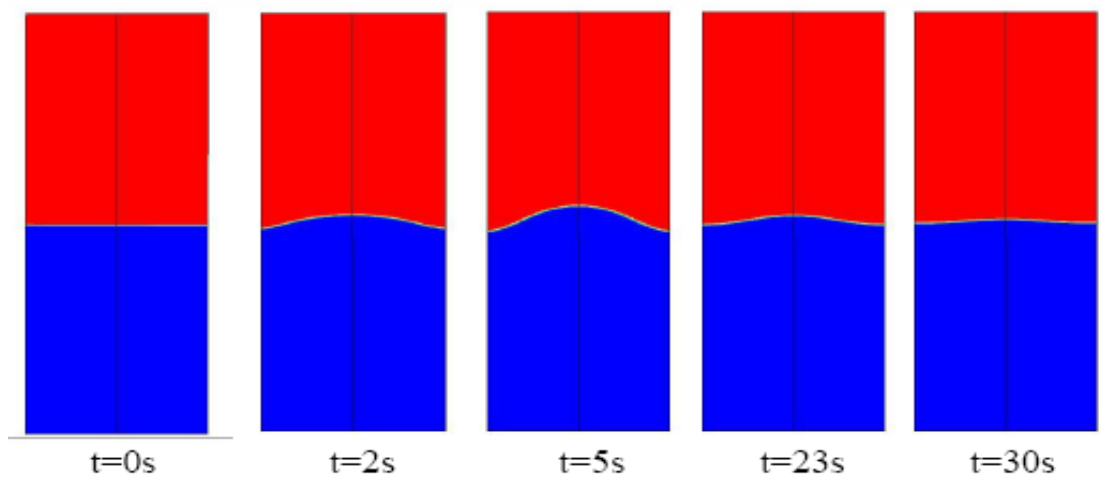

**FIG. 3:** The shape of the oil-water interface at different time moments (Re=74.9, ω=10.6 rad/sec).

In Fig.3, the blue color shows the area occupied by water, and the red color shows the area occupied by silicone oil at different time moments when the cylinder starts rotating at a speed of ω=10.6 rad/sec (Re=74.9). The maximum displacement h of the interface of liquids in the calculations is observed on the axis and for the parameters Re = 74.9, ω =10.6 rad/sec and $R_v$ = 1 at the fifth second, after which, oscillating in time, it slowly levels out.

Figure 4 shows the dependence of the maximum vertical displacement *h* of the interface on the initial plane horizontal surface. The graphs in Fig. 4 show a good correspondence of the calculations with the experiment [1, 2].

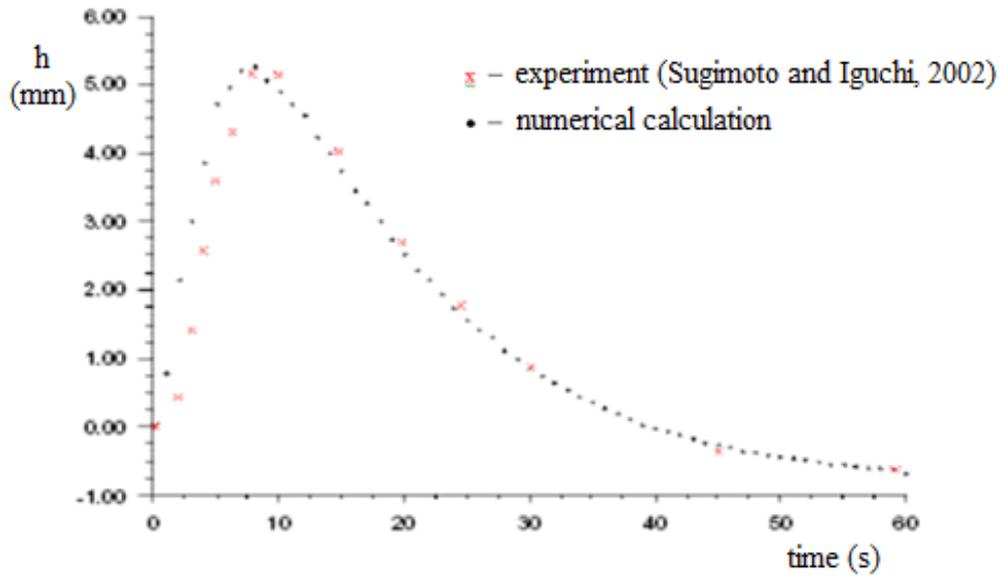

**FIG. 4:** Time variation of the vertical displacement h of the silicone oil–water interface (Re = 74.9).

Calculations were carried out until of the physical time equal to 60 seconds. Animated images of interface behavior and fluid flow were obtained. The fluid flow is complex with multi-vortex and oscillatory, with thin boundary layers at the walls of the cylinder and near the interface. Due to the different properties of liquids, the flows structures of silicone oil and water are different. This can be seen, for example, in Fig.5, where fields of density, velocity, stream function, tracks and isobars are shown for the time t= 5 sec. This figure shows a thin flow structure near the interface, a different flow structure in water and oil, as well as a different pressure distribution in silicone oil and water.

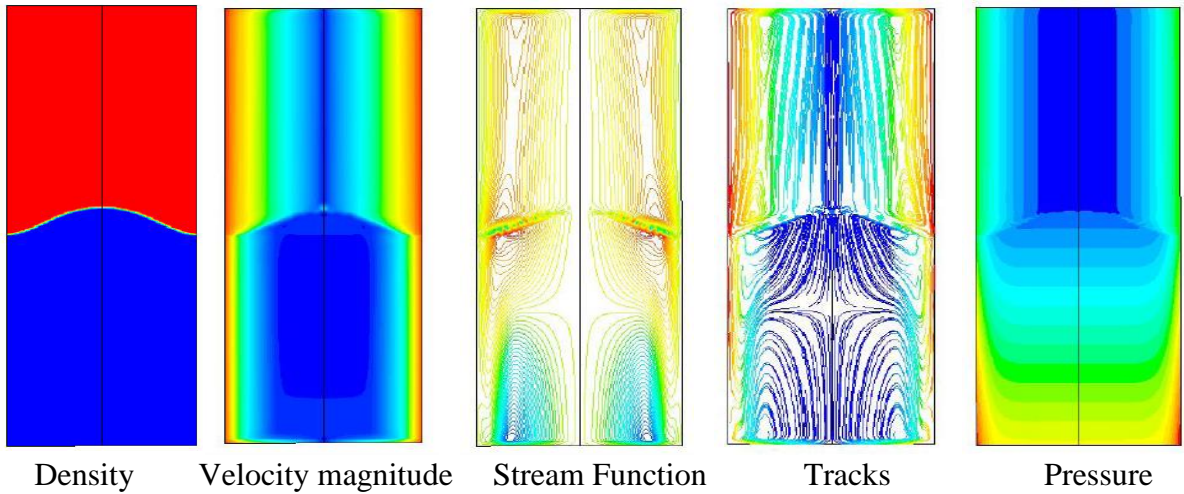

| Density | Velocity magnitude | Stream Function | Tracks | Pressure |

**FIG. 5:** The distributions of density, velocity, stream function, tracks and total pressure at the moment (time=5 sec) of the maximum deviation of the interface from the horizontal position for Re = 74.9

**Conclusions**

Benchmark of mathematical model using a VOF-CSF method for the numerical simulation unsteady shape of the interphase boundary for silicon oil-water systems are performed. Data on the structure of the flow caused by rotation, the maximum altitude of lifting of interface and its shape are presented. A comparison with the experiment data of the results of numerical simulation of a non-stationary change of the shape of the interface between two liquids showed good accuracy of the mathematical model and the calculation method.

1. Sugimoto, T. and Iguchi, M., Behavior of Immiscible Two Liquid Layers Contained in Cylindrical Vessel Suddenly Set in Rotation, *ISIJ Int.,* vol. **42**, pp. 338–343, 2002.

2. Sugimoto, T. and Iguchi, M., Rapid Mixing and Separation of Molten Slag and Metal Using Cylindrical and Baffled Vessels Suddenly Set in Rotation ISIJ Int.l, Vol. 43 (2003), No. 12, pp. 1867–1874